\documentclass{PoS}

\usepackage{amssymb,amsmath}
\newcommand{\Dnu}{\mbox{$\Delta \nu$}}

\newcommand{\numax}{\mbox{$\nu_{\rm max}$}}

\newcommand{\teff}{\mbox{$T_{\rm eff}$}}
\newcommand{\logg}{\mbox{$\log g$}}

\title{Precision Stellar Astrophysics in the Kepler Era}

\ShortTitle{Precision Stellar Astrophysics in the Kepler Era}

\author{\speaker{Daniel Huber}%
        \\
       Sydney Institute for Astronomy, School of Physics, University of Sydney, NSW 2006, Australia\\
       E-mail: \email{daniel.huber@sydney.edu.au}}

\abstract{The study of fundamental properties (such as temperatures, radii, masses, and ages) and interior processes (such as convection and angular momentum transport) of stars has implications on various topics in astrophysics, ranging from the evolution of galaxies to understanding exoplanets. In this contribution I will review the basic principles of two key observational methods for constraining fundamental and interior properties of single field stars: the study stellar oscillations (asteroseismology) and optical long-baseline interferometry. I will highlight recent breakthrough discoveries in asteroseismology such as the measurement of core rotation rates in red giants and the characterization of exoplanet systems. I will furthermore comment on the reliability of interferometry as a tool to calibrate indirect methods to estimate fundamental properties, and present a new angular diameter measurement for the exoplanet host star HD\,219134 which demonstrates that diameters for stars which are relatively well resolved ($\gtrsim 1$\,mas for the $K$ band) are consistent across different instruments. Finally I will discuss the synergy between asteroseismology and interferometry to test asteroseismic scaling relations, and give a brief outlook on the expected impact of space-based missions such as K2, TESS and Gaia.}

\FullConference{Frank N. Bash Symposium 2015\\
		 18-20 October\\
		 The University of Texas at Austin, USA}

\begin{document}

\section{Introduction and Motivation}

The Cosmic Popularity Ladder \cite{drake01}, which measures the ``relative visceral appeal'' of different fields in astrophysics, lists stars in second to last place just above studies of the Sun. Yet, nearly every field in astrophysics is in one way or another dependent on stellar models (which themselves, of course, rely heavily on heliophysics). For example, understanding the evolution of galaxies relies on models of simple stellar populations, which in turn depend on the input physics for stellar isochrones \cite{conroy09}. On a smaller scale, properties of exoplanets depend on the characteristics of host stars, and in many cases the uncertainties on planet radii and masses are dominated by the uncertainties on the properties of the host star \cite{rowe14}. Evidently, improving our understanding of stars in various stages of their evolution is important to advance the field of astrophysics as a whole.

\begin{figure}
\begin{center}
\resizebox{13cm}{!}{\includegraphics{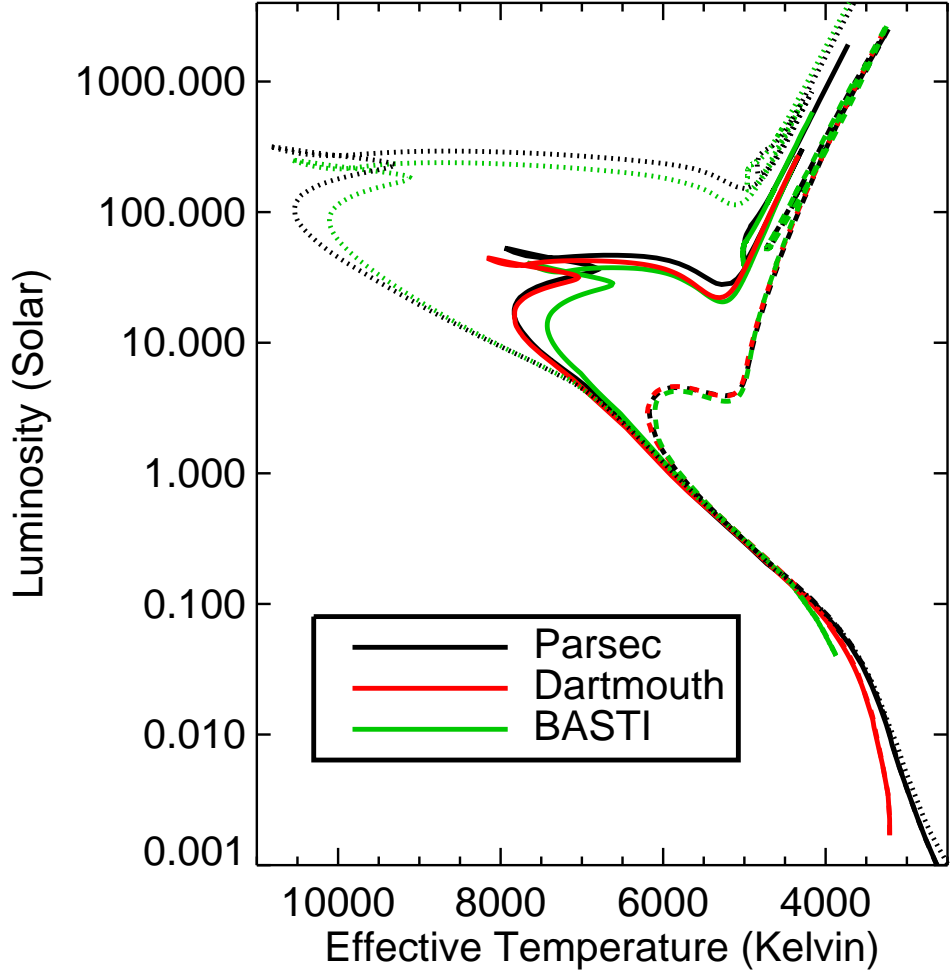}}
\caption{Isochrones with solar chemical composition and ages of 0.1 (dotted lines), 1 (solid lines) and 3 (dashed lines) Gyr taken from the Dartmouth (red), Parsec (black) and BASTI (green) databases. Differences in the models for high and low mass stars are mostly due to variations in poorly constrained input physics.}
\label{iso}
\end{center}
\end{figure}

Despite numerous advances over the last decades, many problems in stellar interior physics remain unsolved. Figure \ref{iso} compares isochrones of different ages from the Dartmouth \cite{dotter08}, Parsec \cite{bressan12} and BASTI \cite{basti} databases with solar composition in an H-R diagram. As expected, the models agree well for stars similar to the Sun. However, significant differences arise for cooler and hotter stars: for example, uncertainties in the description of convective core overshooting lead to different main-sequence lifetimes for intermediate mass stars, while uncertainties in the treatment of convection lead to different predictions of radii for low-mass dwarfs. For red giants, major uncertainties include interior angular momentum transport and mass loss, which can lead to differences in inferred ages of up to 50\% even if other stellar properties are well constrained \cite{casagrande14}.

Over the last decade, space-based photometry has revolutionized our understanding of stars through the application of asteroseismology - the study of stellar pulsations - across the H-R diagram. At the same time, technical advances in long-baseline interferometry have enabled direct measurements of fundamental stellar properties for hundreds of stars. In this contribution I will review the basic principles of asteroseismology and interferometry, and highlight recent advances as well as synergies between both techniques for the study of single field stars. Finally, I will give a brief outlook on the expected results from future observations.

\section{Asteroseismology}

\subsection{Basic Principles of Asteroseismology}

\begin{figure}
\begin{center}
\resizebox{14cm}{!}{\includegraphics{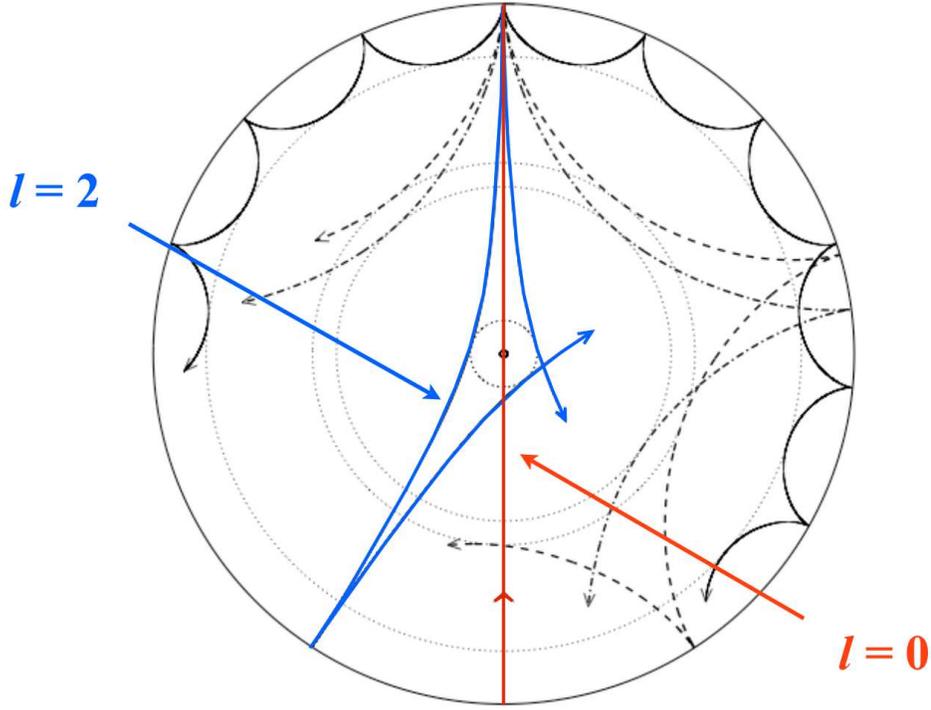}}
\caption{Schematic cross-section of a star illustrating the paths 
of modes with different spherical degrees. Red and blue lines highlight a radial ($l=0$) and quadrupole ($l=2$) mode, respectively. Adapted from \cite{CD03}.}
\label{sphharm}
\end{center}
\end{figure}

Stellar pulsations are observed across the H-R diagram and can be broadly divided into driving mechanisms due to opacity changes in the interior (so-called classical pulsators such as $\delta$\,Scuti stars and Cepheids) and stars which oscillate due turbulent surface convection (so-called solar-like oscillators). In both cases, oscillations (``modes'') can be described with spherical harmonics of degree $l$ (the total number of node lines on the surface), azimuthal order $|m|$ (the number of node lines crossing the equator), and radial order $n$ (the number of nodes from the surface to the center). Radial pulsations are hence denoted as $l=0$, while $l>0$ denote non-radial pulsations. Oscillations with larger spherical degrees penetrate to shallower depths within the star (Figure \ref{sphharm}).

Oscillations can furthermore classified into pressure modes (p modes) and gravity modes (g modes). Pressure modes are acoustic waves, with the pressure gradient acting as the restoring force. Gravity modes are pulsations due to the balance of buoyancy and gravity, with buoyancy acting as the restoring force. Gravity modes are damped in convection zones, and therefore usually only propagate in the interior for cool stars. Pressure modes propagate mostly in radiative zones, and hence are more easily excited with amplitudes that are observable at the surface.

Frequencies of high $n$ and low $l$ can be described by the asymptotic theory of stellar oscillations \cite{vandakurov68,tassoul80,gough86}, which predicts a series of regularly spaced modes. The large frequency separation \Dnu\ is the separation of modes of the same spherical degree $l$ and consecutive radial order $n$, and can be shown to be equal to the inverse of the sound travel time through the stellar diameter \cite{ulrich86,CD03}:

\begin{equation}
\Delta\nu = \left(2 \int^{R}_{0} \frac{dr}{c} \right)^{-1} \: ,
\label{seismo:dnu_phys}
\end{equation}

\noindent
where $c$ is the sound speed. For adiabacity and an ideal gas 
$c \propto\sqrt{T/\mu}$ and $T\propto \mu M / R$, where $\mu$ is the mean molecular weight. Hence, Equation (\ref{seismo:dnu_phys}) can be expressed as \cite{KB95}:

\begin{equation}
\Delta\nu \propto \left(\frac{M}{R^3}\right)^{1/2} \: .
\label{equ:delnu}
\end{equation}

\noindent
The large separation is hence directly proportional to the mean density of a star.
Small frequency separations, which are differences of modes with different degree $l$ and same order $n$, are sensitive to the sound-speed gradient in the stellar interior, and hence the chemical composition during stellar evolution and therefore stellar age. This can be  understood by the fact that modes of different $l$ travel to different depths within the star (Figure \ref{sphharm}), and hence their frequency differences provide information about the radial structure of the star. 

An additional observable is the frequency of maximum power, \numax. The frequency of maximum power has been suggested to scale with the acoustic cut-off frequency \cite{brown91,belkacem11}, which is the upper limit for the reflection of an acoustic 
mode \cite{CD03}:

\begin{equation}
\nu_{\rm ac} = \frac{c}{2 H_{\rm p}} \: .
\end{equation}

\noindent
For an isothermal atmosphere the pressure scale height is $H_{\rm p} = \frac{P R^2}{G M \rho}$. Hence, combined with the ideal gas equation, we can express \numax\ as:

\begin{equation}
\nu_{\rm max} \propto \nu_{\rm ac} \propto \frac{M}{R^2 \sqrt{T_{\rm eff}}} \: .
\label{equ:numax}
\end{equation}

\noindent
Measurements of \numax, \Dnu, or individual frequencies provide a powerful way to determine the stellar structure and fundamental properties 
(such as radius and mass), which are otherwise difficult to determine for field stars.

\subsection{The Space Photometry Revolution of Asteroseismology}

\begin{figure}
\begin{center}
\resizebox{11cm}{!}{\includegraphics{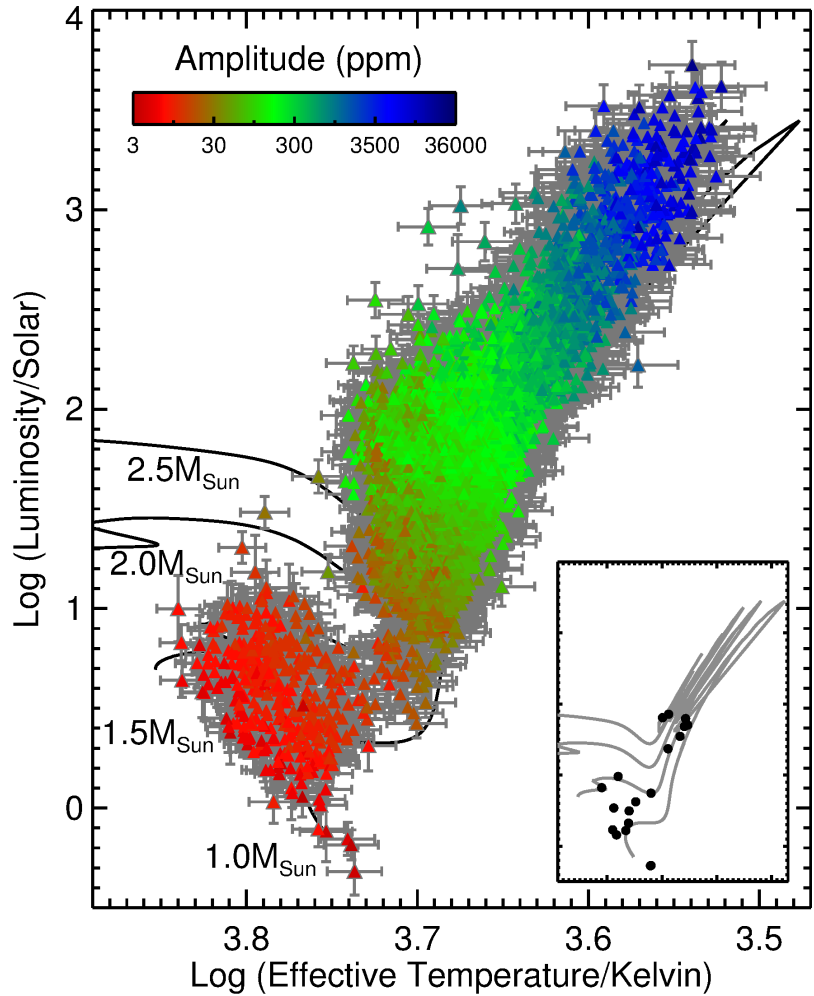}}
\caption{H-R diagram showing stars with detected solar-like oscillations by \textit{Kepler} \cite{huber11,stello13,huber14}. Color coding marks the mean amplitude per radial mode of the observed oscillations. Solid lines show solar metallicity evolutionary tracks with different masses as indicated in the plot. The inset shows the same diagram with all detections circa 2008 (prior to the launch of CoRoT). }
\label{seismohrd}
\end{center}
\end{figure}

Early asteroseismic observations were mostly focused on radial velocity campaigns \cite{brown91,kjeldsen95,bedding01,carrier01,bouchy01,frandsen02,butler04}, leveraging on the improved measurements of Doppler velocities for the detection of exoplanets. The first space-based photometric observations obtained by the Canadian MOST telescope led to detections in a few red giants \cite{barban07,kallinger08b} and Procyon \cite{guenther08,huber11}, and further space-based observations were performed using the WIRE (Wide-Field Infrared Explorer) startracker \cite{schou01,retter03,bruntt05,stello08}, the SMEI (Solar Mass Ejection Imager) satellite \cite{tarrant07} and the Hubble Space Telescope \cite{edmonds96,gilliland08,stello09b,gilliland11}. In total, observations prior to 2009 yielded detections in $\sim 20$ stars (see inset of Figure \ref{seismohrd}). 

A major breakthrough, which is commonly referred to as the beginning of the space photometry revolution of asteroseismology, was achieved by the CoRoT (Convection Rotation and Planetary Transits) satellite. CoRoT detected oscillations in a number of main sequence stars \cite{michel08} and several thousands red giant stars \cite{hekker09,mosser11b}, demonstrating for the first time that red giants oscillate in non-radial modes \cite{deridder09}. This opened the door for detailed studies of the interior structure of red giants which were previously not thought to be possible.

The \textit{Kepler} space telescope continued the revolution of cool-star asteroseismology by covering the low-mass H-R diagram with detections, including 500 dwarfs and subgiants \cite{chaplin11a} as well as over 15,000 red giants \cite{hekker11c,stello13}. The large number of oscillating red giants is predominantly due to the fact oscillation amplitudes increase with luminosity (Figure \ref{seismohrd}), therefore biasing asteroseismic detections towards more evolved stars. Additionally, the 30-minute sampling cadence for most \textit{Kepler} targets sets a limit of $\logg \lesssim 3.5$, since less evolved stars oscillate above the Nyquist frequency. Overall, the sample of cool stars amenable for asteroseismology has grown by two orders of magnitude in the last 8 years.

\subsection{Probing the Interior of Red Giants}

While for main sequence stars the propagation cavities of p modes and g modes are well separated, the different core and envelope densities in evolved stars can lead to evolutionary stages where p modes and g modes have similar frequencies. This gives rise to mixed modes \cite{dziembowski01}, which contain contributions from g modes in the core but unlike pure g modes have low enough mode inertias to be observable at the surface. While high order p modes are equally spaced in frequency, g modes are predicted to be equally spaced in period. The coupling of p modes with g modes causes mixed modes to be shifted from their original frequency spacing \cite{aizenman77}, yielding multiple frequencies per radial order which are expected to be approximately equally spaced in period. 

\begin{figure*}
\begin{center}
\resizebox{\hsize}{!}{\includegraphics{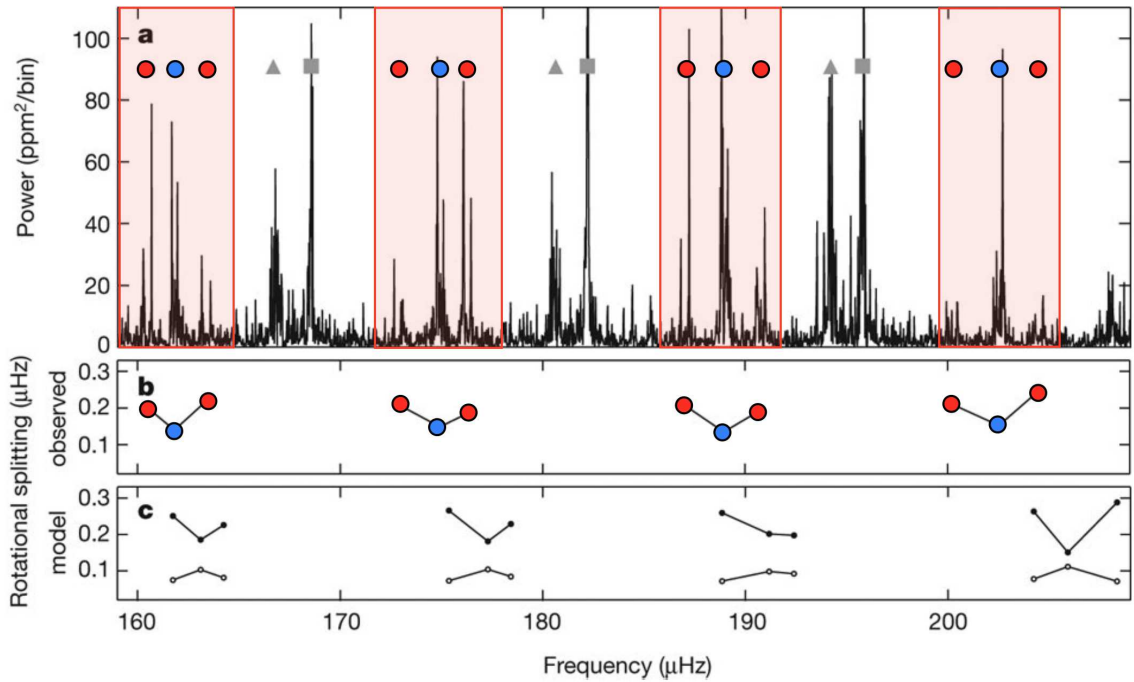}}
\caption{Top panel: Power spectrum of a red giant observed by \textit{Kepler}. Dipole ($l=1$) modes are highlighted in red and marked by red and blue circles, while radial ($l=0$) and quadrupole ($l=2$) modes are marked by squares and triangles, respectively. There are multiple $l=1$ modes per radial order due to the coupling of p modes with g modes in the core (mixed modes). Middle panel: Measured rotational splitting of each mixed $l=1$ mode in the top panel. The mixed modes in the wings of each radial order (highlighted as red circles), which are more sensitive to the core, show a larger rotational splitting and hence faster rotation than modes in the center of each order (highlighted as blue circles). Bottom panel: Rotational splittings according to a model with a core spinning 10 times faster than the envelope (top) and with rigid rotation (bottom). Adapted from \cite{beck12}.}
\label{fig:rotation}
\end{center}
\end{figure*}

The detection of mixed $l=1$ modes in red giants observed by \textit{Kepler} \cite{bedding10} led to a series of groundbreaking discoveries in our understanding of the interiors of red giants. Following the detection of equal period spacings \cite{beck11}, it was demonstrated that giants ascending the RGB and He-core burning red giants can be separated based on their mixed-mode period spacing \cite{bedding11,mosser11b}. Shortly after, it was discovered that mixed modes are split into multiplets by rotation, and that frequency splittings for g-dominated mixed modes are substantially higher than for p-dominated mixed modes due radial differential rotation \cite{beck12}. Figure \ref{fig:rotation} shows a typical power spectrum of a red giant exhibiting mixed modes with the signature of radial differential rotation.

\textit{Kepler} data has since allowed measurements of core rotation rates for hundreds of red giants, allowing an unprecedented view into the internal rotation evolution of evolved stars \cite{mosser13,deheuvels14}. The data show that the cores spin up as stars evolve towards the RGB, followed by a gradual spin-down as stars evolve towards the He-core burning main sequence. Predicted core rotation rates in models are up to factors of 10--100 larger than observed \cite{marques13,cantiello14,fuller14}, pointing to a yet unidentified mechanism responsible for transporting angular momentum from the core to the envelope. Finally, in the latest twist in red giant asteroseismology, the puzzling absence of dipole modes observed in some red giants \cite{mosser11c} has been explained by strong internal magnetic fields which can trap oscillations in the radiative core, and hence can be used to place upper limits on the interior magnetic field strengths \cite{fuller15,stello16}. These remarkable observational insights by asteroseismology, ranging from interior composition, rotation to magnetic fields, promise to enable important theoretical advances in our understanding of the interior structure of red giant stars for years to come.

\subsection{Asteroseismology of Exoplanet Host Stars}

In addition to testing stellar models, asteroseismology has recently started to play an important role the study of exoplanets \cite{huber15b}. This synergy has largely been enabled by the fact that high-precision time domain observations (either in intensity or velocity) can be simultaneously used to detect exoplanets and study stellar oscillations.

To date there are approximately 100 exoplanet host stars for which oscillations have been detected, most of which have been observed by \textit{Kepler} \cite{huber13}. The primary application of asteroseismology is to measure host star radii, which combined with transit depths yield planet radii to a precision of $\sim$\,1--2\% in the best cases \cite{howell12,ballard14}. Asteroseismology has been used to measure the radius of the smallest exoplanet known to date \cite{barclay12} and determine precise ages of dozens of exoplanets \cite{silva15}, including the oldest known terrestrial-sized planets with an age of $\sim$\,11\,Gyr \cite{campante15}.

\begin{figure}[t!]
\begin{center}
\resizebox{13cm}{!}{\includegraphics{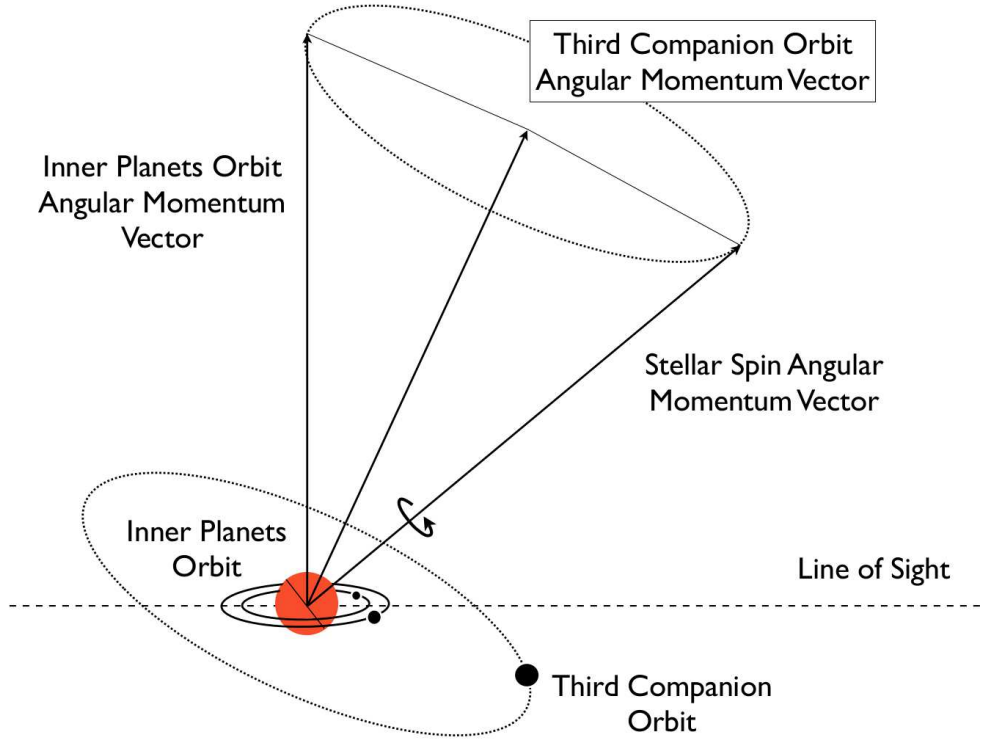}}
\caption{Graphical sketch of the Kepler-56 system. The spin-axis inclination of the red giant host star was determined from asteroseismology, while the inner transiting planets were detected through transits. The torque of the outer companion, which was detected through radial-velocity follow-up observations, causes a precession of the orbital axis of the inner transiting planets and the stellar spin axis. Both precessions occur 
at different rates, causing a periodic spin-orbit misalignment. Sizes are not to scale. From \cite{huber13b}.}
\label{kepler56}
\end{center}
\end{figure}

Another remarkable application has been the use of asteroseismology to measure the spin-axis inclination of host stars through the relative heights of the $l(l+1)$ modes in rotationally split multiplets \cite{gizon03}. Stellar inclinations are important for studying the architecture and dynamical history of transiting exoplanets by constraining the angle between the stellar spin axis and the axis of the planetary orbit (the obliquity). Since the presence of transits shows that the orbital axis is perpendicular to the line of sight, a low stellar inclination automatically implies a misalignment of the orbital plane and the equatorial plane of the star (a high obliquity). 

\textit{Kepler} and CoRoT have enabled asteroseismic inclination measurements for several exoplanet systems \cite{chaplin13c,gizon13,benomar14,vaneylen14,lund14}. One of the most intriguing examples is Kepler-56, a red giant hosting two transiting planets. The Kepler-56 power spectrum shows dipole modes which are split into triplets, yielding an inclination of $47\pm6$ degrees and demonstrating the first stellar spin-orbit misalignment in a multiplanet system \cite{huber13b}. Radial velocity observations revealed a long-term trend due to a massive companion on a wide orbit which, under the assumption of a significant mutual inclination to the inner transiting planets, can explain the misalignment through a precession of the orbital axis of the inner planets due to the torque of the wide companion (Figure \ref{kepler56}) \cite{li14}. This scenario has previously been proposed theoretically \cite{mardling10,kaib11,batygin12}, and demonstrated for the first time that spin-orbit misalignments are not confined to hot Jupiter systems.

In addition to precise exoplanet radii and orbital architectures, asteroseismology can also be used to constrain orbital eccentricities \cite{sliski14,vaneylen15}. The variety of  applications highlights the importance of stellar astrophysics for the field of exoplanet science, and the need to improve stellar models in order to advance our understanding of exoplanets.

\section{Optical Long-Baseline Interferometry}

\subsection{Basic Principles of Interferometry}

\begin{figure}
\begin{center}
\resizebox{13cm}{!}{\fbox{\includegraphics{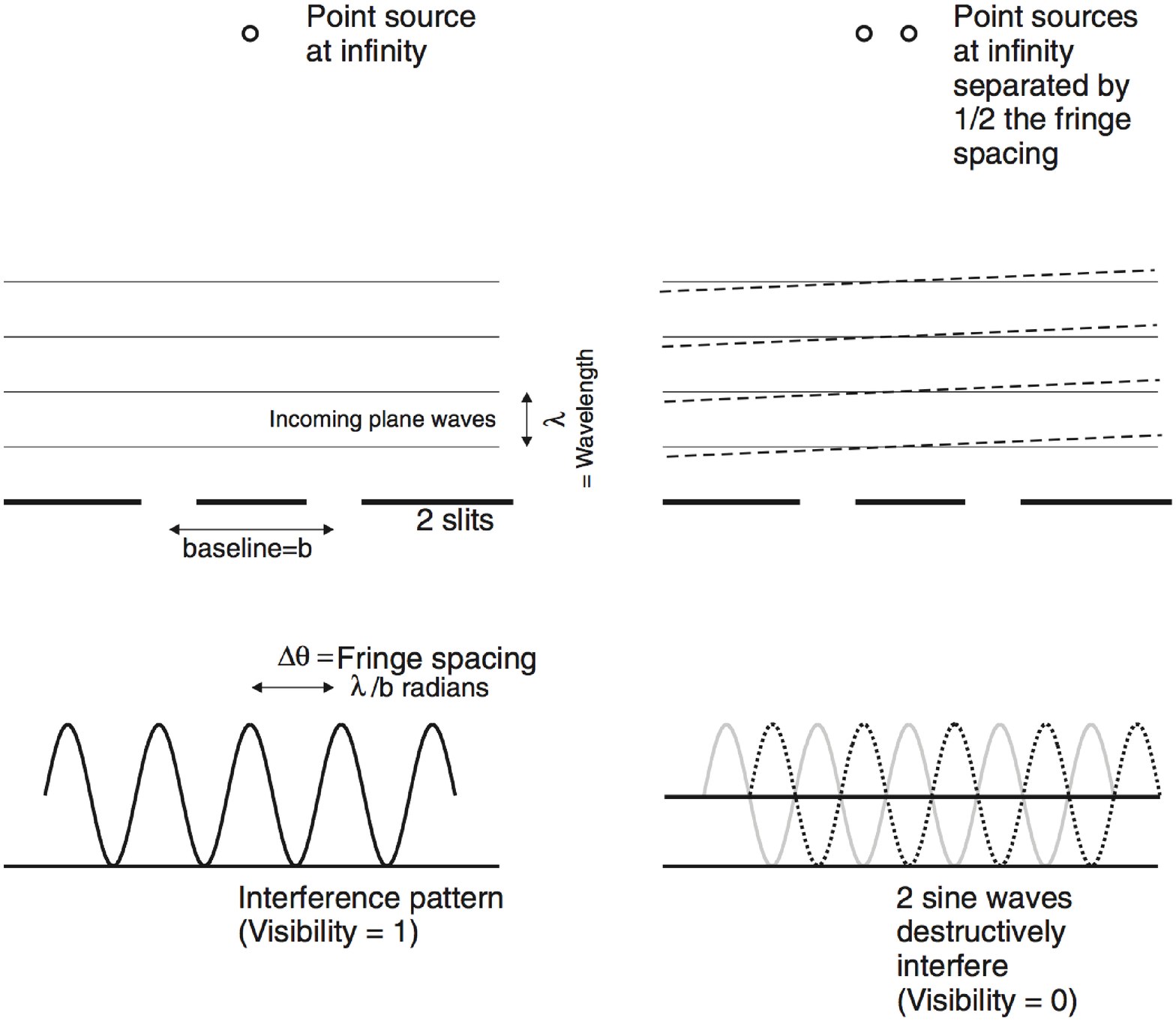}}}
\caption{Young's two slit experiment illustrating the basic principle of stellar 
interferometry. From \cite{monnier03}.}
\label{young}
\end{center}
\end{figure}

The principle of interferometry is traditionally illustrated by a variation of Young's double slit experiment performed in the early 19th century (Figure \ref{young}). As monochromatic light from a point source passes through a double slit (left panel), an interference (``fringe'') pattern is formed. Each point of the wavefront is a source of spherical wavelets, which constructively and destructively interfere at the screen behind the slits. The spacing between dark and bright patches of the interference pattern is:

\begin{equation}
\Delta\Theta = \frac{\lambda}{b} \: ,
\end{equation}

\noindent
where $\lambda$ is the wavelength and $b$ the separation between the slits (the ``baseline''). Suppose that the light hitting the double slit is emitted by two point sources that are separated in angle by half the fringe spacing ($\frac{\lambda}{2b}$) 
(right panel of Figure \ref{young}). In this case, the fringes are in anti-phase and the interference patterns form an incoherent sum to give an evenly illuminated screen. 

It is evident from this simple discussion that any intermediate separation of the 
two point sources or a single extended source will result in different contrasts. This fringe contrast is the ratio of the fringe amplitude to the fringe intensity:

\begin{equation}
V = \frac{I_{\rm max}-I_{\rm min}}{I_{\rm max}+I_{\rm min}} \: ,
\label{equ:vis}
\end{equation}

\noindent
where $V$ is the fringe visibility. A measurement of the visibility at a given wavelength and separation of the two slits (or telescopes) is therefore directly related to the structure of the object being observed. Stating the above more generally, it can be shown that the complex visibility is equal to the spatial Fourier transform of the intensity distribution of a given object \cite{monnier03}. Therefore, measuring visibilities at different baselines in principle allows image reconstruction with angular resolutions that far exceed the diffraction limit of the largest telescopes in the world.

\subsection{Stellar Angular Diameters and the CHARA Array}

The most basic measurement in stellar interferometry is the angular size of a star. The total energy radiated by a star per unit surface area and per unit time is: 

\begin{equation}
F = \sigma T_{\rm eff}^{4} \: ,
\label{equ:sb2}
\end{equation}

\noindent
where $\sigma$ is the Stefan-Boltzmann constant and $T_{\rm eff}$ is the effective temperature. Given a star with a radius $R$ at a distance $d$, the total (bolometric) flux emitted over all wavelengths $f_{\rm bol}$ received by an observer on Earth is:

\begin{equation}
f_{\rm bol} = F \frac{R^2}{d^2} \; .
\label{equ:bolflux}
\end{equation}

\noindent
Combined with the straightforward relation $R = d \frac{\theta}{2}$ for the apparent angular diameter $\theta$, Equation (\ref{equ:sb2}) can be rearranged as:

\begin{equation}
T_{\rm eff} = \left(\frac{4 f_{\rm bol}}{\sigma \theta^2}\right)^{1/4} \: .
\end{equation}

\noindent
The angular diameter of a star combined with a distance (e.g. from a trigonometric parallax) or with a measurement of the received bolometric flux (e.g. from spectrophotometry) therefore enables a measurement of the radius and 
effective temperature of star. Importantly, both measurements (in particular the radius) are almost model independent. An important exception is the limb darkening correction for the angular diameter, which can become the dominant source of uncertainty if the visibility curve is well constrained (see next section).

The first stellar angular diameter measurement dates back to Michelson \cite{michelson}, who visually determined the size of $\alpha$\,Ori using a 20-foot interferometer mounted on the 100-inch Hooker telescope on Mt.\ Wilson observatory. The first systematic study was performed by Hanbury Brown and collaborators \cite{hanbury74} using the Narrabri Intensity Interferometer to measure diameters of 32 bright stars, which subsequently led to the first calibration of color-temperature relationships \cite{code}. 

\begin{figure}
\begin{center}
\resizebox{15cm}{!}{\includegraphics{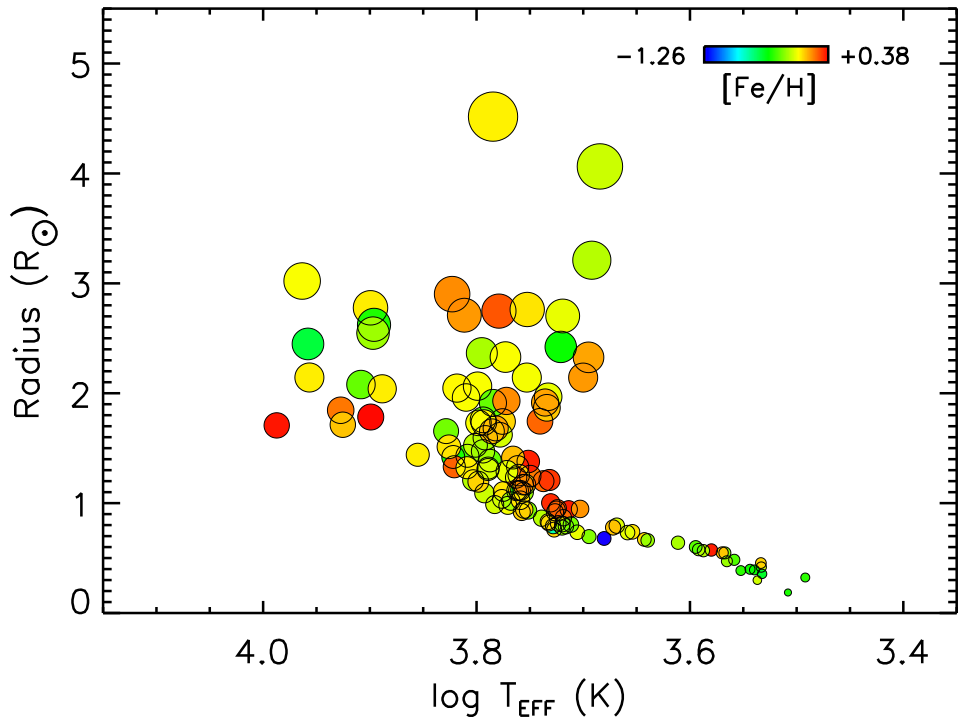}}
\caption{Stellar radius versus effective temperature derived from angular diameters measurements with the CHARA Array. The symbol size scales with the size of the star, and the color coding denotes the spectroscopically determined metallicity. From \cite{boyajian13}.}
\label{charahrd}
\end{center}
\end{figure}

The following decades saw the development of a number of modern interferometers, most notably the CHARA (Center for High Angular Resolution Astronomy) Array \cite{brummelaar04} located at Mount Wilson Observatory. The array consists of six 1-m class telescopes arranged in a non-redundant Y-shaped configuration with baselines ranging from 34\,m to a maximum of 331\,m. The CHARA array uses instruments in both optical and near-infrared wavelengths, and is currently the largest operating optical long-baseline interferometer in the world. Since commencing science operations in 2004 it has produced several breakthrough results, such as the first image of a single main-sequence star \cite{monnier07}. One of the most influential results of the CHARA Array in terms of fundamental stellar properties have been angular diameter measurements of more than a hundred stars across the H-R diagram \cite{boyajian12,boyajian12b,boyajian13}, including several exoplanet host stars \cite{baines09,ligi12,vonbraun14}, yielding empirical radii and temperatures to test stellar atmosphere and interior models across a range of spectral types and evolutionary states (Figure \ref{charahrd}). 

\subsection{Calibrating Models and Indirect Methods: How Accurate is Interferometry?}

Interferometric angular diameters have traditionally been used as benchmarks to test stellar models and calibrate indirect methods to estimate fundamental properties of stars. For example, CHARA diameters have confirmed that interior models overpredict radii of M dwarfs by up to 10\% for a fixed \teff\ \cite{boyajian12b}, in agreement with constraints from eclipsing binary systems \cite{carter11,kraus11}. CHARA diameters have since been used to calibrate methods to predict radii of late-type dwarfs for which direct measurements are not possible \cite{mann15,newton15} and (in combination with eclipsing binaries) explore new physical mechanisms to improve models for cool stars \cite{feiden13,feiden14,macdonald14}. Such calibrations are particularly important for exoplanet transit surveys such as K2 \cite{howell14} and TESS \cite{ricker14}, which predominantly target late-type dwarfs.

While interferometry is often considered as the ``ground-truth'', it is important to realize that interferometric data can be affected by strong systematic errors. Observed visibilities are reduced by atmospheric turbulence and hence need to be calibrated by observing unresolved point sources as closely in time and position on the sky as possible. In practice calibrators are often somewhat resolved, and hence systematic errors in the assumed calibrator sizes propagate into uncertainties of the angular diameter, in particular if the target is not well resolved \cite{vanbelle05}. Additional uncertainties include errors in the wavelength scale and limb-darkening correction. However, such error sources are not always taken into account, leading to uncertainties that sometimes fail to encompass differences between diameter measurements by different groups and instruments. 

Such differences can have a significant impact on the calibration of more indirect methods. A recent comparison of effective temperatures determined from the infrared flux method (IRFM) with interferometry revealed a systematic difference for CHARA $K$-band diameters to predict higher temperatures for stars with angular sizes $\lesssim$\,1\,mas \cite{casagrande14c}, while smaller diameters measured in $H$ band showed better agreement \cite{huang15b}. Since calibration errors are more severe for smaller diameters (corresponding to more unresolved sources, given a fixed baseline and wavelength), this indicates that some diameters measured with CHARA may be affected by systematic errors.

\begin{figure}
\begin{center}
\resizebox{13cm}{!}{\includegraphics{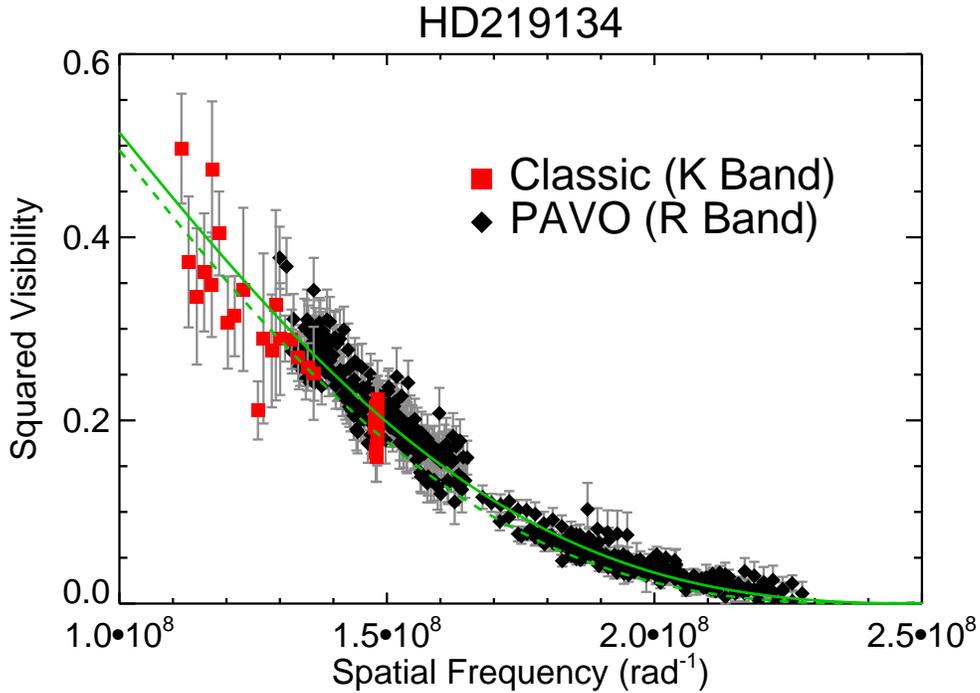}}
\caption{Squared visibility verus spatial frequency for the exoplanet host star HD\,219134 measured in the $K$ band using the Classic beam combiner (red squares) \cite{boyajian12} and preliminary new data obtained with the $R$ band PAVO beam combiner (black diamonds). The green solid line shows the fitted diameter model to the PAVO data, while the dashed green line shows the fit to the Classic data.}
\label{fig:hd219134}
\end{center}
\end{figure}

To test the accuracy of interferometric angular diameters, we have recently started cross-calibration efforts with the CHARA Array by revisiting stars which were observed with the Classic ($H/K$ band) beam combiner with the higher-resolution PAVO ($R$ band) beam combiner \cite{ireland08}. Figure \ref{fig:hd219134} shows preliminary results for the K dwarf HD\,219134, the closest transiting exoplanet known to date \cite{montalebi15,vogt15}. PAVO data were taken over four nights in August and September 2015 using two baselines (E2W2, W1W2), and calibrated using HD\,218376 ($\theta=0.304$\,mas), HD\,225289 ($\theta=0.206$\,mas) and HD\,223386 ($\theta=0.177$\,mas). The PAVO data resolve HD\,219134 nearly down to the first null, making them largely independent of calibrator diameter uncertainties. The Classic (red squares) and PAVO (black diamonds) data show good agreement, with limb-darkened diameters of $\theta_{\rm{LD,Classic}}=1.093\pm0.012$\,mas (using $\mu_{K}=0.29\pm0.05$) and $\theta_{\rm{LD,PAVO}}=1.109\pm0.008$\,mas (using $\mu_{R}=0.67\pm0.05$). Limb darkening coefficients were taken from \cite{claret10} for a solar metallicity, $\teff=4750$\,K and $\logg=4.5$ model, and a 5\% and 3\,nm uncertainty was assumed for the calibrator diameters and wavelength scale, respectively. Note that the PAVO diameter is dominated by the uncertainty in the limb-darkening correction, while the Classic diameter is dominated by measurement/calibrator diameter uncertainties\footnote{Note that the Classic uncertainty is larger than the published value due to the inclusion of uncertainties in calibrator sizes, limb darkening and wavelength scale. The uniform disc diameters are $\theta_{\rm{UD,Classic}}=1.066\pm0.011$\,mas and $\theta_{\rm{UD,PAVO}}=1.033\pm0.005$\,mas, respectively.}. Future observations will focus on measuring fringes in the second visibility lobe, which will allow direct constraints on the limb-darkening in the $R$ band.

While the results for HD\,219134 demonstrate that $\gtrsim$1\,mas diameters measured in $K$ band are likely reliable, cross-calibrations for smaller diameters and for a larger number of beam combiners are needed. Some results have shown good agreement for stars as small as 0.6\,mas (16\,Cyg\,A) \cite{white13}, while other measurements (such as 18\,Sco and $\theta$\,Cyg) have shown stark discrepancies \cite{bazot11,white13}. In addition to revisiting diameters, we are also pursuing simultaneous dual beam combiner observations of the same target and calibrators with baselines corresponding to near-identical spatial frequencies for each instrument. This will eliminate most sources of systematic errors except for the absolute visibility calibration between instruments, which is still poorly understood.

\subsection{Combining Interferometry and Asteroseismology}

The combination of asteroseismology and interferometry allows a powerful tests of interior models. Using asteroseismic densities and interferometric angular diameters, near-model independent measurements of the radius, mass and effective temperature can be made for single stars, and hence directly compared to evolutionary models. Such approaches have been used to confirm that 18\,Sco is a true solar twin \cite{bazot11}, detect a potential discrepancy in interior models for the metal-rich \textit{Kepler} star HD\,173701 \cite{huber12b}, and constrain the modeling of individual oscillation frequencies for solar-type stars \cite{metcalfe12,gruberbauer13,metcalfe15}.

\begin{figure}[t!]
\begin{center}
\resizebox{13cm}{!}{\includegraphics{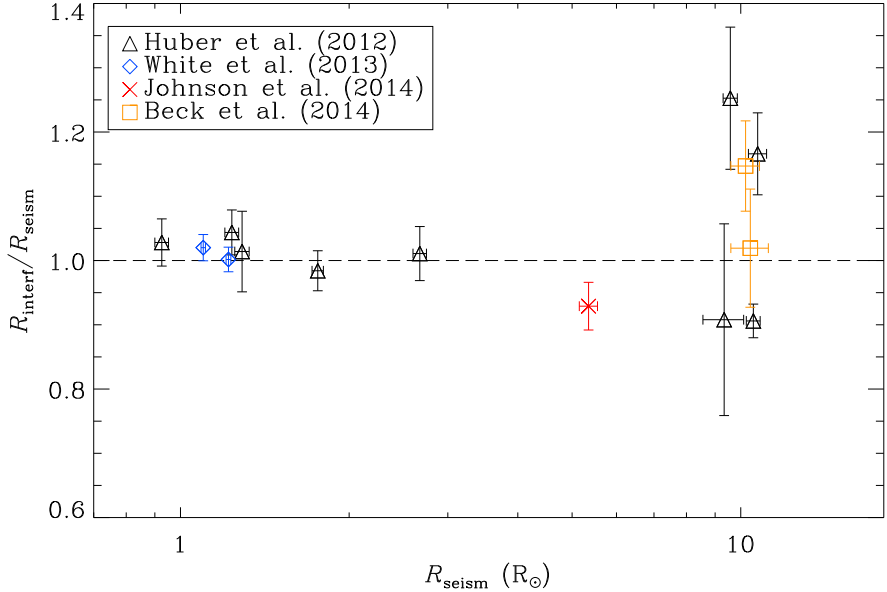}}
\caption{Interferometric radii divided by asteroseismic radii derived from scaling relations as a function of asteroseismic radii. Different colors and symbols denote the source of the measurements as given in the legend. From \cite{white15}.}
\label{testseismo}
\end{center}
\end{figure}

Alternatively, interferometry can also be used to test asteroseismology. In particular, asteroseismic scaling relations (Equations \ref{equ:delnu} and \ref{equ:numax}) have become increasingly popular to determine stellar properties. However, these relations are only approximate and require careful calibration, in particular for stars which are significantly more evolved than the Sun. Figure \ref{testseismo} shows comparions between radii from both methods based on a collection of high S/N asteroseismic detections and well resolved diameters \cite{huber12b,white13,johnson14,beck15}. The comparison shows excellent agreement between interferometric radii and asteroseismic radii for dwarfs with a scatter of $\lesssim 4\%$, while giants are dominated by large uncertainties in the Hipparcos parallaxes. Future observations will alleviate this problem by focusing on brighter red giants, as well as the use of more precise parallaxes which will soon be available from the Gaia satellite.

\section{Future Prospects}

\begin{figure}
\begin{center}
\resizebox{12cm}{!}{\includegraphics{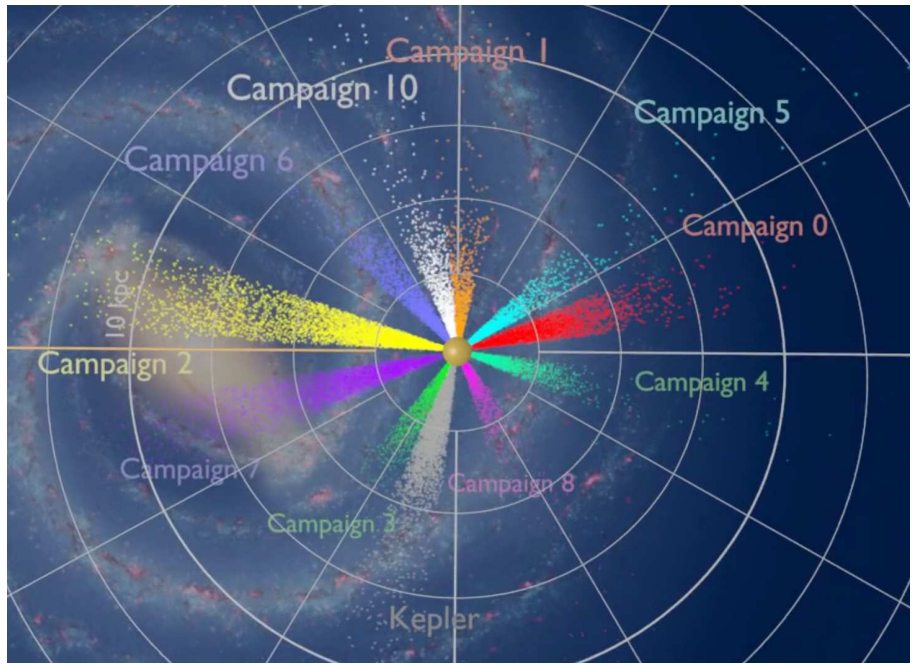}}
\caption{Illustration of the Milky Way with predicted detections of oscillating red giants targeted in the first 10 campaigns of the K2 Mission (colored dots). Asteroseismology combined with spectroscopy will allow to map the metallicity and age distribution of stars in an unprecedented manner. Image Credit: Kristin Riebe (Potsdam), Sanjib Sharma (Sydney) and Dennis Stello (Sydney).}
\label{gap}
\end{center}
\end{figure}

Asteroseismology and interferometry will continue to flourish as tools for stellar astrophysics over the coming decades. A particularly important synergy concerns the calibration of asteroseismic scaling relations for giants. While space-based missions will continue the asteroseismic revolution initiated by \textit{Kepler}, ground-based surveys such as APOGEE \cite{apokasc}, GALAH \cite{desilva15} and LAMOST \cite{zhao06} are obtaining spectra of thousands of stars throughout the galaxy. Recognizing the powerful synergy between both methods, major efforts are currently underway to combine asteroseismic and spectroscopic data to probe the chemo-dynamical history of stellar populations in our Galaxy (often referred to as ``galactic archeology''). In particular, Kepler's follow-up mission K2 \cite{howell14} is measuring oscillations of red giants throughout the ecliptic plane \cite{stello15,huber16}, therefore dramatically expanding the galactic field of view of \textit{Kepler} and CoRoT (see Figure \ref{gap}). However, the determination of stellar ages depends sensitively on stellar masses, and hence asteroseismic scaling relations. Therefore, the success of galactic archeology relies crucially on our ability to accurately calibrate asteroseismology using direct methods such as interferometry. 

Importantly, future space-based missions will also obtain asteroseismic observations of stars which will be significantly brighter than typical \textit{Kepler} targets, and hence amenable to interferometric follow-up observations. At the same time, the increased sensitivity of the CHARA Array with an upcoming adaptive-optics upgrade will further increase the overlap between stellar samples that can be observed with both techniques. Combined with the release of parallaxes by Gaia, the sample of single stars with directly measured fundamental properties will allow unprecedented tests of stellar models over the coming decade. 

\vspace{0.25cm}

\noindent
\textbf{Acknowledgements:} I thank Frank Bash, Rachael Livermore, Stefano Meschiari and all our hosts at UT Austin for a fantastic and memorable meeting. Financial support was provided by the Australian Research Council's Discovery Projects funding scheme (project number DE140101364).

\bibliographystyle{JHEP}
\bibliography{../tex/references.bib}


\end{document}